\documentclass[a4paper,12pt]{article}

\usepackage{latexsym}
\usepackage[cp866]{inputenc}
\usepackage[english]{babel}

\usepackage{graphicx}

\input epsf

\begin{document}

\title{Nature of light scalars through photon-photon
collisions\thanks{The plenary talk given by N.N. Achasov at the
International Workshop on $e^+e^-$ Collisions from Phi to Psi
(PhiPsi09), October 13-16, 2009, IHEP, Beijing, China}}

\author{N.N. Achasov and G.N. Shestakov
\\[0.1cm] \small {\it Laboratory of Theoretical Physics, Sobolev Institute
for Mathematics,}\\ \small {\it Novosibirsk, 630090, Russia}}

\date{}
\maketitle

\vspace{-2mm}
\begin{abstract}
The surprising thing is that arising almost 50 years ago from the
linear sigma model (LSM) with spontaneously broken chiral symmetry,
the light scalar meson problem has become central in the
nonperturbative quantum chromodynamics (QCD) for it has been made
clear that LSM could be the low energy realization of QCD. First we
review briefly signs of  four-quark nature of  light scalars. Then
we show that  the light scalars are produced in the two photon
collisions via four-quark transitions in contrast to the classic $P$
wave tensor $q\bar q$ mesons that are produced via two-quark
transitions $\gamma\gamma\to q\bar q$. Thus we get new evidence of
the four-quark nature of these states.
\end{abstract}

\vspace{4mm} \noindent{\bf\large 1\ \ Introduction} \vspace{2mm}

The scalar channels in the region up to 1 GeV became a stumbling
block of QCD. The point is that both perturbation theory and sum
rules do not work in these channels because there are not solitary
resonances in this region. At the same time the question on the
nature of the light scalar mesons is major for  understanding the
mechanism of the chiral symmetry realization, arising from the
confinement, and hence for understanding the confinement itself.

\vspace{4mm} \noindent{\bf\large 2\ \ QCD, chiral limit,
confinement, \boldmath{$\sigma$}-models} \vspace{2mm}

$L=-(1/2)Tr\left (G_{\mu\nu}(x)G^{\mu\nu}(x)\right )+\bar
q(x)(i\hat{D}-M)q(x).$ $M$ mixes left and right spaces $q_L(x)$ and
$q_R(x)$. But in chiral limit $M\to 0$ these spaces separate
realizing $U_L(3)\times U_R(3)$ flavour symmetry. Experiment
suggests, confinement forms colourless observable hadronic fields
and spontaneous breaking of chiral symmetry with massless
pseudoscalar fields. There are two possible scenarios for QCD at low
energy. 1. $U_L(3)\times U_R(3)$ non-linear $\sigma$-model. 2.
$U_L(3)\times U_R(3)$ linear $\sigma$-model. The experimental nonet
of the light scalar mesons suggests $U_L(3)\times U_R(3)$ linear
$\sigma$-model.

\vspace{4mm} \noindent{\bf\large 3\ \ History of light scalar
mesons} \vspace{2mm}

Hunting the light $\sigma$ and $\kappa$ mesons had begun in the
sixties already. But long-standing unsuccessful attempts to prove
their existence in a conclusive way entailed general disappointment
and an information on these states disappeared from PDG Reviews. One
of principal reasons against the $\sigma$ and $\kappa$ mesons was
the fact that both $\pi\pi$ and $\pi K$ scattering phase shifts do
not pass over $90^0$ at putative resonance masses. [Meanwhile, there
were discovered the narrow light scalar resonances, the isovector
$a_0(980)$ and isoscalar $f_0(980)$.]

\vspace{4mm} \noindent{\bf\large 4\ \ \boldmath $SU_L(2)\times
SU_R(2)$ linear $\sigma$-model \cite{AS94}} \vspace{2mm}

Situation changes when we showed that in the linear $\sigma$-model
$$ L=\frac{1}{2}\left
[(\partial_\mu\sigma)^2+(\partial_\mu\overrightarrow{\pi})^2\right
]- \frac{m_\sigma^2}{2}\sigma^2- \frac{m_\pi^2}{2}
\overrightarrow{\pi}^2 - \frac{m_\sigma^2-m_\pi^2}{8f^2_\pi}\left
[\left (\sigma^2+\overrightarrow{\pi}^2\right )^2 +4f_\pi\sigma
\left (\sigma^2+\overrightarrow{\pi}^2\right ) \right ]^2 $$ there
is a negative background phase which hides the $\sigma$ meson. It
has been made clear that shielding wide lightest scalar mesons in
chiral dynamics is very natural. This idea was picked up and
triggered new wave of theoretical and experimental searches for the
$\sigma$ and $\kappa$ mesons.

\vspace{4mm} \noindent{\bf\large 5\ \ Our approximation
\cite{AS94,AS07}} \vspace{2mm}

Our approximation is as follows (see Fig. 1):
$$ T_0^{0(tree)}=\frac{m_\pi^2-m_\sigma^2}{32\pi f^2_\pi}\left
[5-3\frac{m_\sigma^2-m_\pi^2}{m_\sigma^2-s}-2\frac{m_\sigma^2-
m_\pi^2}{s-4m_\pi^2}\ln\left(1+\frac{s-4m^2_\pi}{m_\sigma^2}\right
)\right], $$
$$ T^0_0=\frac{T_0^{0(tree)}}{1-i\rho_{\pi\pi}T_0^{0(tree)}}=
\frac{e^{2i\left(\delta_{res}+\delta_{bg}\right)}-1}{2i\rho_{\pi\pi}}=
\frac{e^{2i\delta^0_0}-1}{2i\rho_{\pi\pi}}=T_{bg}+
e^{2i\delta_{bg}}T_{res}\,, $$
$$ T_{bg}=\frac{e^{2i\delta_{bg}}-1}{2i\rho_{\pi\pi}}\,, \ \
T_{res}=\frac{e^{2i\delta_{res}}-1}{2i\rho_{\pi\pi}}\,, \ \
\rho_{\pi\pi}=\sqrt{1-\frac{4m_\pi^2}{s}}\,, $$
$$ T_{bg}=\frac{\lambda (s)}{1-i\rho_{\pi\pi}\lambda(s)}\,, \ \
\lambda(s)=\frac{m_\pi^2-m_\sigma^2}{32\pi f^2_\pi}\left
[5-2\frac{m_\sigma^2-m_\pi^2}{s-4m_\pi^2}\ln\left
(1+\frac{s-4m^2_\pi}{m_\sigma^2}\right)\right]\,, $$
$$ T_{res}=\frac{1}{\rho_{\pi\pi}}.\frac{\sqrt{s}\Gamma_{res}(s)}{M^2_{res}
-s+\mbox{Re}\Pi_{res}(M^2_{res})-\Pi_{res}(s)}\,, \ \ M^2_{res}=
m_\sigma^2 - \mbox{Re}\Pi_{res}(M^2_{res})\,,$$
$$ \mbox{Im}\Pi_{res}(s)=\sqrt{s}\Gamma_{res}(s)=\frac{g_{res}^2(s)}{16\pi}
\rho_{\pi\pi}\,, \ \
\mbox{Re}\Pi_{res}(s)=-\frac{g_{res}^2(s)}{16\pi}\lambda(s)
\rho_{\pi\pi}^2\,, $$
$$ g_{res}(s)=\frac{g_{\sigma\pi\pi}}{\bigl
|1-i\rho_{\pi\pi}\lambda(s)\bigr |}\,, \ \
g_{\sigma\pi\pi}=\sqrt{\frac{3}{2}}\,g_{\sigma\pi^+\pi^-}\,, \ \
g_{\sigma\pi^+\pi^-}=\frac{m^2_\pi-m^2_\sigma}{f_\pi}\,, $$
$$ T^2_0=\frac{T_0^{2(tree)}}{1-i\rho_{\pi\pi}T_2^{0(tree)}}
=\frac{e^{2i\delta_0^2}-1}{2i\rho_{\pi\pi}}\,, $$
$$ T^{2(tree)}_0=\frac{m_\pi^2-m_\sigma^2}{32\pi f^2_\pi}\left
[2-2\frac{m_\sigma^2-m_\pi^2}{s-4m_\pi^2}\ln\left
(1+\frac{s-4m^2_\pi}{m_\sigma^2}\right )\right]\,. $$

\begin{figure}\centerline{\epsfysize=1.5in 
\epsfbox{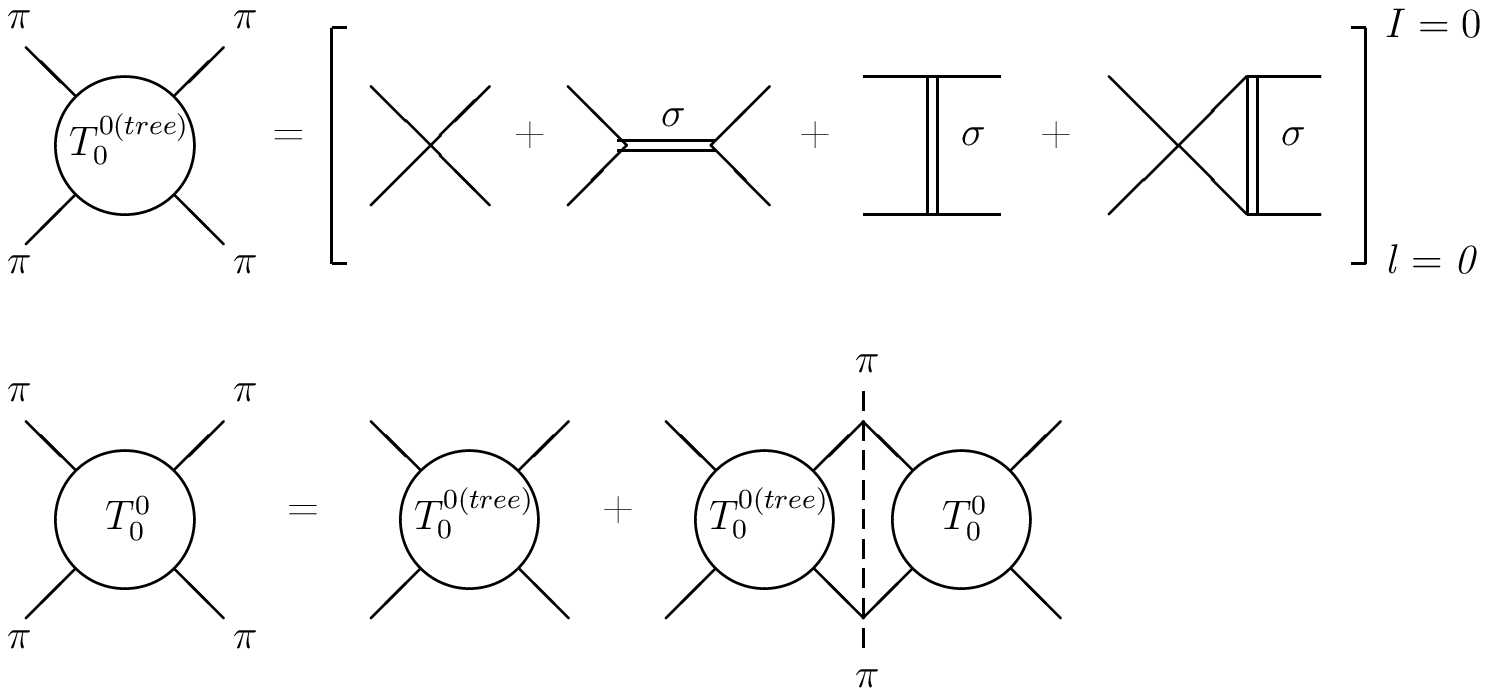}}\vspace{-2mm} \caption{{\footnotesize The
graphical representation of the $S$ wave $I=0$ $\pi\pi$ scattering
amplitude $T^0_0$.}}\end{figure}

\vspace{4mm} \noindent{\bf\large 6\ \ Chiral shielding in \boldmath
$\pi\pi\to\pi\pi$ \cite{AS94,AS07}} \vspace{2mm}

The results in our approximation are: $M_{res}$ =$0.43$\,GeV,
$\Gamma_{res}(M^2_{res})$\,=\,$0.67$\,GeV, $m_\sigma$\,=0.93\,GeV,
$\Gamma^{renorm}_{res}(M^2_{res})$\,=\,$\frac{\Gamma_{res}(
M^2_{res})} {1+d[\mbox{\small Re}\Pi_{res}(s)]/ds|_{s=M^2_{res}}}
$\,=\,0.53\,GeV, $g_{res}(M^2_{res})/g_{\sigma\pi\pi}$=0.33,
$a^0_0$=\,0.18\, $m_\pi^{-1}$, $a^2_0$=$-0.04\, m_\pi^{-1}$, the
Adler zeros $(s_A)^0_0$=$0.45\, m^2_\pi$ and $(s_A)^2_0$=$2.02\,
m^2_\pi$.

The chiral shielding of the $\sigma(600)$ meson in
$\pi\pi$\,$\to$\,$\pi\pi$ is illustrated in Fig. 2 with the help of
the $\pi\pi$ phase shifts $\delta_{res}$, $\delta_{bg}$,
$\delta^0_0=\delta_{res}+\delta_{bg}$ (a), and with the help of the
corresponding cross sections (b).

\begin{figure}\vspace{3mm}\centerline{\epsfysize=1.6in
\epsfbox{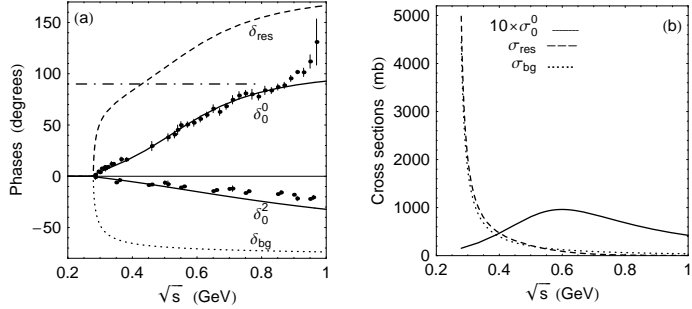}}\vspace{-2mm} \caption {{\footnotesize The
$\sigma$ model. Our approximation.}}\end{figure}

\vspace{4mm} \noindent{\bf\large 7\ \ The \boldmath $\sigma$ pole
and $\sigma$ propagator \cite{AS07}} \vspace{2mm}

In the pole $T^0_0\to g^2_\pi/(s-s_R)$, where
$g^2_\pi=(0.12+i0.21)$\,GeV$^2$, $s_R=(0.21-i0.26)$\,GeV$^2$,
$\sqrt{s_R}=M_R-i\Gamma_R/2=(0.52-i0.25)$\,GeV. Considering the
residue of the $\sigma$ pole in $T^0_0$ as the square of its
coupling constant to the $\pi\pi$ channel is not a clear guide to
understand the $\sigma$ meson nature for its great obscure imaginary
part.

Another matter the $\sigma$ meson propagator
$$ \frac{1}{D_\sigma (s)}= \frac{1}{M^2_{res}-s
+\mbox{Re}\Pi_{res}(M^2_{res})- \Pi_{res}(s)}\,.$$ The $\sigma$
meson self-energy $\Pi_{res}(s)$ is caused by the intermediate
$\pi\pi$ states, that is, by the four-quark intermediate states.
This contribution shifts the Breit-Wigner (BW) mass greatly
$m_\sigma-M_{res}$\,=\,0.50\,GeV. So, half the BW mass is determined
by the four-quark contribution at least. The imaginary part
dominates the propagator modulus in the region
300\,MeV\,$<\sqrt{s}<$\,600\,MeV. So, the $\sigma$ field is
described by its four-quark component at least in this energy
(virtuality) region.

\vspace{4mm} \noindent{\bf\large 8\ \ Chiral shielding in \boldmath
$\gamma\gamma\to\pi\pi$ \cite{AS07}} \vspace{2mm}

The $\gamma\gamma\to\pi^+\pi^-$ reaction amplitude is given by\\
$$T_S(\gamma\gamma\to\pi^+\pi^-)=
T_S^{Born}(\gamma\gamma\to\pi^+\pi^-)+8\alpha I_{\pi^+\pi^-} 
T_S(\pi^+\pi^-\to\pi^+\pi^-)$$ $$=[
T_S^{Born}(\gamma\gamma\to\pi^+\pi^-)+8\alpha I_{\pi^+\pi^-} 
(2T_0^0+T_0^2)/3]\mbox{\ \ in elastic region}$$
$$=\frac{2}{3}e^{i\delta^0_0}
\left\{T_S^{Born}(\gamma\gamma\to\pi^+\pi^-)\cos\delta^0_0+
\frac{8\alpha}{\rho_{\pi\pi}}(\mbox{Re}I_{\pi^+\pi^-})
\sin\delta^0_0\right\}$$
$$+\frac{1}{3}e^{i\delta^2_0}\left\{T_S^{Born}(\gamma\gamma\to\pi^+
\pi^-)\cos\delta^2_0+\frac{8\alpha}{\rho_{\pi\pi}}(
\mbox{Re}I_{\pi^+\pi^-})\sin\delta^2_0\right\}.$$

The $\gamma\gamma\to\pi^0\pi^0$ reaction amplitude is given by
$$T_S(\gamma\gamma\to\pi^0\pi^0)= 8\alpha I_{\pi^+\pi^-}
T_S(\pi^+\pi^-\to\pi^+\pi^-)$$ $$=16\alpha I_{\pi^+\pi^-}
(T_0^0-T_0^2)/3\mbox{\ \ in elastic region}$$ $$=
\frac{2}{3}e^{i\delta^0_0}
\{T_S^{Born}(\gamma\gamma\to\pi^+\pi^-)\cos\delta^0_0+
\frac{8\alpha}{\rho_{\pi\pi}}(\mbox{Re}I_{\pi^+\pi^-})
\sin\delta^0_0\}$$ $$-\frac{2}{3}e^{i\delta^2_0}
\{T_S^{Born}(\gamma\gamma\to\pi^+\pi^-)\cos\delta^2_0+
\frac{8\alpha}{\rho_{\pi\pi}}(
\mbox{Re}I_{\pi^+\pi^-})\sin\delta^2_0\}.$$

Here $T_S^{Born}(\gamma\gamma\to\pi^+\pi^-)=
(8\alpha/\rho_{\pi\pi})\,\mbox{Im}I_{\pi^+\pi^-}\,,$ \
$\alpha$\,=\,1/137, and the triangle $\pi^+\pi^-$ loop integral
$I_{\pi^+\pi^-}=\frac{m^2_\pi}{s}\left
(\pi+i\ln\frac{1+\rho_{\pi\pi}}{1-\rho_{\pi\pi}}\right )^2-1$, for
$s\geq 4m_\pi^2$.

Fig. 3 illustrates the chiral shielding of the $\sigma(600)$ in the
cross sections $\gamma\gamma\to\pi\pi$.

\begin{figure}\begin{center}\begin{tabular}{cc}{\epsfysize=1.6in
\epsfbox{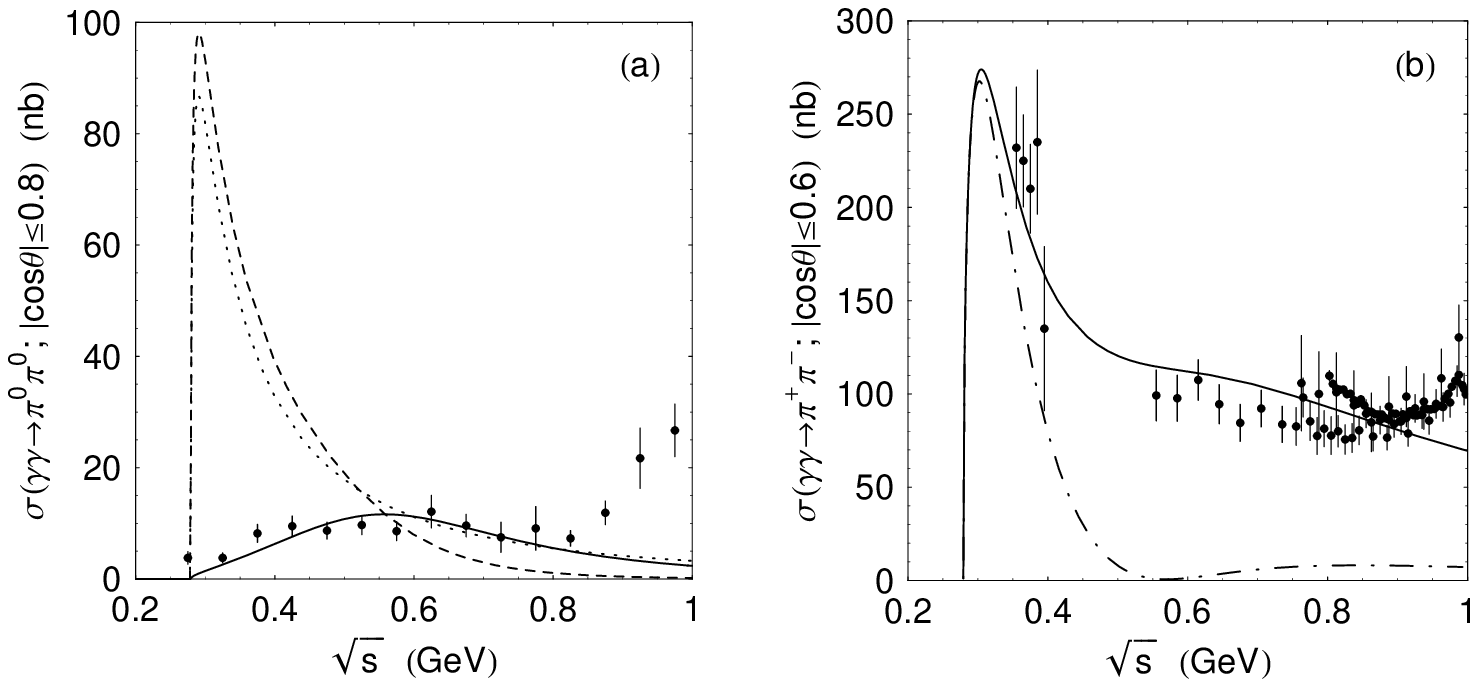}}&
{\epsfysize=1.6in\epsfbox{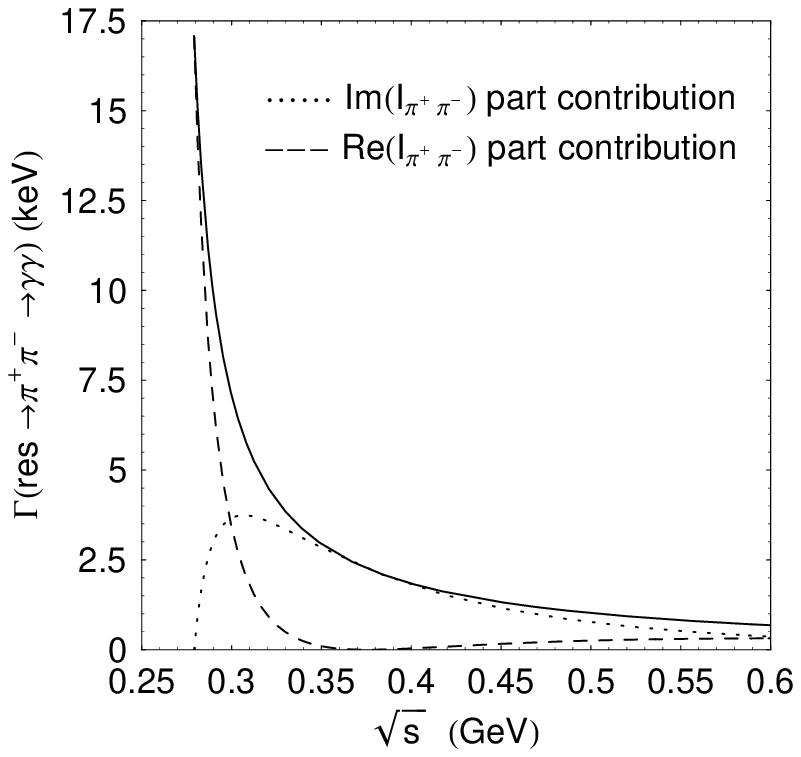}}\\[0.15cm]
{\small Figure 3:}{\footnotesize \ (a) The solid, dashed, and dotted
lines are $\sigma_S(\gamma\gamma\to\pi^0\pi^0)$,}
& {\small Figure 4:}{\footnotesize \ The $\sigma\to\gamma\gamma$}\\
{\footnotesize $\sigma_{res}(\gamma\gamma\to\pi^0\pi^0)$, and
$\sigma_{bg}(\gamma\gamma\to\pi^0\pi^0)$. (b) The dashed-dotted line
is} & {\footnotesize decay width.}\\
{\footnotesize $\sigma_S(\gamma\gamma\to\pi^+\pi^-)$. The solid line
includes the higher waves from} & \\ {\footnotesize
$T^{Born}(\gamma\gamma\to\pi^+\pi^-)$.} &
\end{tabular}\end{center}\end{figure}

\vspace{4mm} \noindent{\bf\large 9\ \ Four-quark transition
\boldmath $\sigma\to\gamma\gamma$ \cite{AS07}} \vspace{2mm}

The energy dependent $\sigma\to\gamma\gamma$ decay width
$$\Gamma(\sigma\to\pi^+\pi^-\to\gamma\gamma,s)=
\frac{1}{16\pi\sqrt{s}}\left|g(\sigma\to\pi^+\pi^-\to\gamma\gamma,s
)\right|^2,$$ where $g(\sigma\to\pi^+\pi^-\to\gamma\gamma,\,s)=
(\alpha/2\pi)\,I_{\pi^+\pi^-}\,g_{res\,\pi^+\pi^-}(s)$, is shown in
Fig. 4.

So, the the $\sigma\to\gamma\gamma$ decay is described by the
triangle $\pi^+\pi^-$ loop diagram $res\to\pi^+\pi^-\to\gamma\gamma$
($I_{\pi^+\pi^-}$). Consequently, it is due to the four-quark
transition because we imply a low energy realization of the
two-flavour QCD by means of the the $SU_L(2)\times SU_R(2)$ linear
$\sigma$-model. As the Fig. 4 suggests, the real intermediate
$\pi^+\pi^-$ state dominates in $g(res\to\pi^+\pi^-\to\gamma\gamma)$
in the $\sigma$ region $\sqrt{s}<$\,0.6\,GeV. Thus the picture in
the physical region is clear and informative. But, what  about the
pole in the complex $s$-plane? Does the pole residue reveal the
$\sigma$ indeed?

In the $\sigma$ pole for $\gamma\gamma\to\pi\pi$ one has
$$\frac{1}{16\pi}\sqrt{\frac{3}{2}}\,T_S(\gamma\gamma\to\pi^0\pi^0)\to
\frac{g_\gamma g_\pi}{(s-s_R)}\,,$$ $g_\gamma g_\pi=(-0.45 -
i0.19)\times 10^{-3}\,\mbox{GeV}^2$, $g_\gamma/g_\pi=
(-1.61+i1.21)\times 10^{-3}$,
$\Gamma(\sigma\to\gamma\gamma)=|g_\gamma |^2/M_R\approx
2\,\mbox{keV}$. It is hard to believe that anybody could learn the
complex but physically clear dynamics of the $\sigma\to\gamma\gamma$
decay described above from the residues of the $\sigma$ pole.

\vspace{4mm} \noindent{\bf\large 10\ \ First lessons} \vspace{2mm}

1. Leutwyler and collaborators \cite{CCL06} obtained $$\sqrt{s_R}=
M_R-i\Gamma_R/2 =( 441^{+16}_{-8}-i272^{+12.5}_{-9})\,\mbox{MeV} $$
with the help of the Roy equation. Our result agrees with the above
qualitatively $$ \sqrt{s_R}=M_R-i\Gamma_R/2
=(518-i250)\,\mbox{MeV}\,.$$

2. Could the above scenario incorporates the primary lightest scalar
Jaffe four-quark state \cite{Ja77}? Certainly the direct coupling of
this state to $\gamma\gamma$ via neutral vector pairs
($\rho^0\rho^0$ and $\omega\omega$), contained in its wave function,
is negligible $\Gamma(q^2\bar
q^2\to\rho^0\rho^0+\omega\omega\to\gamma\gamma) \approx
10^{-3}$\,keV as we showed in 1982 \cite{ADS82}. But its coupling to
$\pi\pi$ is strong and leads to $\Gamma(q^2\bar
q^2\to\pi^+\pi^-\to\gamma\gamma)$ similar to
$\Gamma(res\to\pi^+\pi^-\to\gamma\gamma)$ in the above
Fig.~\ref{fig4}. Let us add  to $T_S(\gamma\gamma\to\pi^0\pi^0)$ the
amplitude for the the direct coupling of $\sigma$ to $\gamma\gamma$
conserving unitarity
$$
T_{direct}(\gamma\gamma\to\pi^0\pi^0)=sg^{(0)}_{\sigma\gamma\gamma}
g_{res}(s)e^{i\delta_{bg}}/D_{res}(s)\,,
$$
where $g^{(0)}_{\sigma\gamma\gamma}$ is the direct coupling constant
of $\sigma$ to $\gamma\gamma$, the factor $s$ is caused by gauge
invariance. Fitting the $\gamma\gamma\to\pi^0\pi^0$ data gives a
negligible value of $g^{(0)}_{\sigma\gamma\gamma}$,
$\Gamma^{(0)}_{\sigma\gamma\gamma}=
|M^2_{res}g^{(0)}_{\sigma\gamma\gamma}|^2/(16\pi M_{res} )\approx
0.0034$\,keV, in astonishing agreement with our old prediction
\cite{ADS82}.

3. The majority of current investigations of the mass spectra in
scalar channels does not study particle production mechanisms. That
is why such investigations are only preprocessing experiments, and
the derivable information is very relative. The only progress in
understanding the particle production mechanisms could essentially
advance us in revealing the light scalar meson nature, as  is
evident from the foregoing.

\vspace{4mm} \noindent{\bf\large 11\ \ Troubles and expectancies}
\vspace{2mm}

In theory the principal problem is impossibility to use the linear
$\sigma$-model in the tree level approximation inserting widths into
$\sigma$ meson propagators because such an approach breaks both the
unitarity and the Adler self-consistency conditions. The comparison
with the experiment requires the non-perturbative calculation of the
process amplitudes. Nevertheless, now there are the possibilities to
estimate odds of the $U_L(3)\times U_R(3)$ linear $\sigma$-model to
underlie physics of light scalar mesons in phenomenology, taking
into account the idea of chiral shielding, our treatment of
$\sigma(600)$-$f_0(980)$ mixing based on quantum field theory ideas,
and Adler's conditions \cite{AK906}.

An example of the phenomenological treatment is shown in Fig.~5 with
$g_{\sigma\pi^+\pi^-}^2/4\pi $\,=\,0.99\,GeV$^2$, $g_{\sigma
K^+K^-}^2/4\pi$\,=\,2$ \cdot10^{-4}$\,GeV$^2$,
$g_{f_0\pi^+\pi^-}^2/4\pi $\,=\,0.12\,GeV$^2$, $g_{f_0
K^+K^-}^2/4\pi$ =\,1.04\,GeV$^2$, $m_\sigma$\,=\,679\,MeV,
$\Gamma_\sigma$\,=\,498\,MeV, $m_{f_0}$\,=\,989\,MeV, and the
$J$\,=\,$I$\,=\,0 $\pi\pi$ scattering length
$a^0_0$\,=\,0.223\,$m^{-1}_{\pi^+}$.

\begin{center}
\includegraphics[width=3in,height=1.5in]{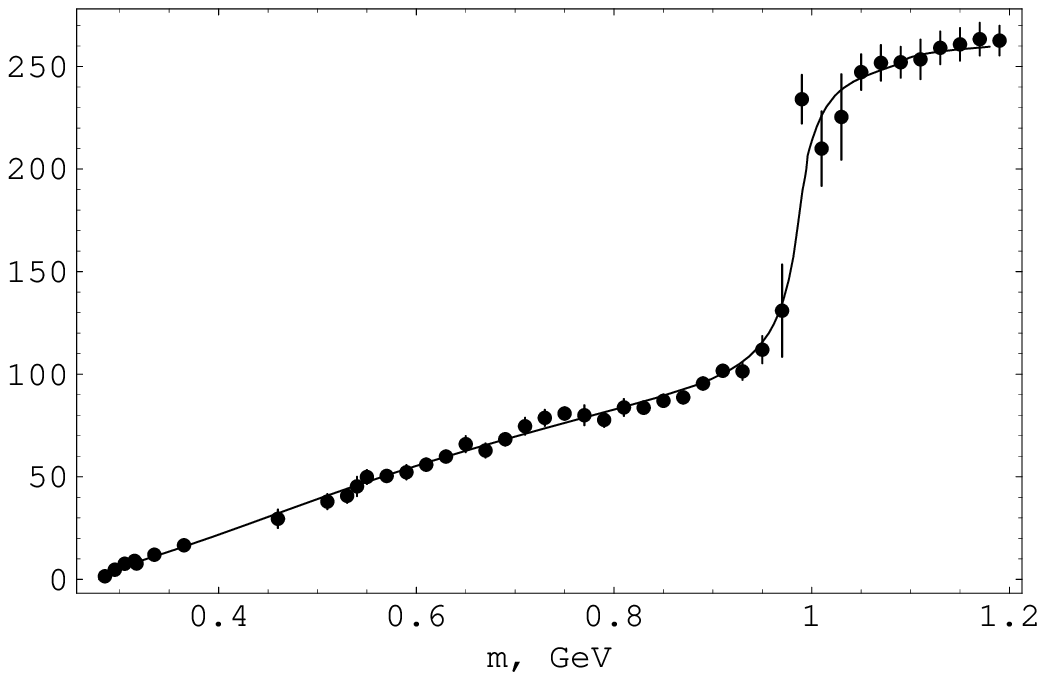}
\\[0.3cm]
{\small Figure 5:}{\footnotesize \ The $\pi\pi$ phase shift
$\delta_0^0=\delta_B^{\pi\pi}+\delta_{res}$.} \end{center}

\vspace{4mm} \noindent{\bf\large 12\ \ Four-quark model}
\vspace{2mm}

The nontrivial nature of the well-established light scalar
resonances $f_0(980)$ and $a_0(980)$ is no longer denied practically
anybody. As for the nonet as a whole, even a cursory look at PDG
Review gives an idea of the four-quark structure of the light scalar
meson nonet, $\sigma (600)$, $\kappa (800)$, $f_0(980)$, and
$a_0(980)$, inverted in comparison with the classical $P$ wave
$q\bar q$ tensor meson nonet $f_2(1270)$, $a_2(1320)$,
$K_2^\ast(1420)$, $f_2^\prime (1525)$. Really, while the scalar
nonet cannot be treated as the $P$ wave $q\bar q$ nonet in the naive
quark model, it can be easy understood as the $q^2\bar q^2$ nonet,
where $\sigma$ has no strange quarks, $\kappa$ has the $s$ quark,
$f_0$ and $a_0$ have the $s\bar s$ pair. Similar states were found
by Jaffe in 1977 in the MIT bag \cite{Ja77}.

\vspace{4mm} \noindent{\bf\large 13\ \ Radiative decays of
\boldmath{$\phi$} meson \cite{A8907}} \vspace{2mm}

Ten years later we showed that $\phi\to\gamma a_0\to\gamma\pi\eta$
and $\phi\to\gamma f_0\to \gamma\pi\pi$ can shed light on the
problem of $a_0(980)$ and $f_0(980)$ mesons. Now these decays are
studied not only theoretically but also experimentally. The
measurements (1998, 2000) were reported by SND and CMD-2. After
(2002) they were studied  by KLOE in agreement with the Novosibirsk
data but with a considerably smaller error. Note that $a_0(980)$  is
produced in the radiative $\phi$ meson decay as intensively as
$\eta'(958)$  containing $\approx 66\% $ of $s\bar s$, responsible
for $\phi\approx s\bar s\to\gamma s\bar s\to\gamma \eta'(958)$. It
is a clear qualitative argument for the presence of the $s\bar s$
pair in the isovector $a_0(980)$ state, i.e., for its four-quark
nature.

\vspace{4mm} \noindent{\bf\large 14\ \ \boldmath{$K^+K^-$} loop
model \cite{A8907}} \vspace{2mm}

When basing the experimental investigations, we suggested one-loop
model $\phi\to K^+K^-\to\gamma a_0/f_0$, see Fig. 6. This model is
used in the data treatment and is ratified by experiment, see Fig.
7. Gauge invariance gives the conclusive arguments in favor of the
$K^+K^-$ loop transition as the principal mechanism of $a_0(980)$
and $f_0(980)$ meson production in the $\phi$ radiative decays.

\begin{center}\begin{tabular}{ccc}
\includegraphics[width=0.9in]{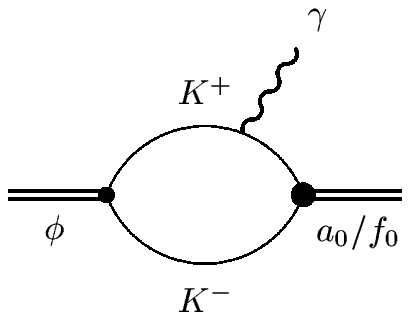}\ \ \ &
\raisebox{-3.0mm}{$\includegraphics[width=0.9in]{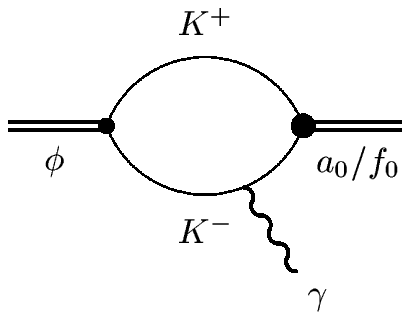}$}\ \ \ &
\includegraphics[width=0.9in]{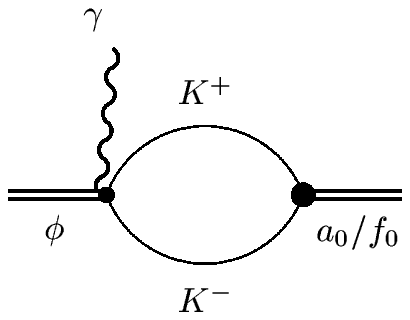}\end{tabular}
\\[0.4cm] {\small Figure 6:}{\footnotesize \ The $K^+K^-$ loop model.}
\end{center}

\begin{center}
\begin{tabular}{cc}
\includegraphics[height=1.6in]{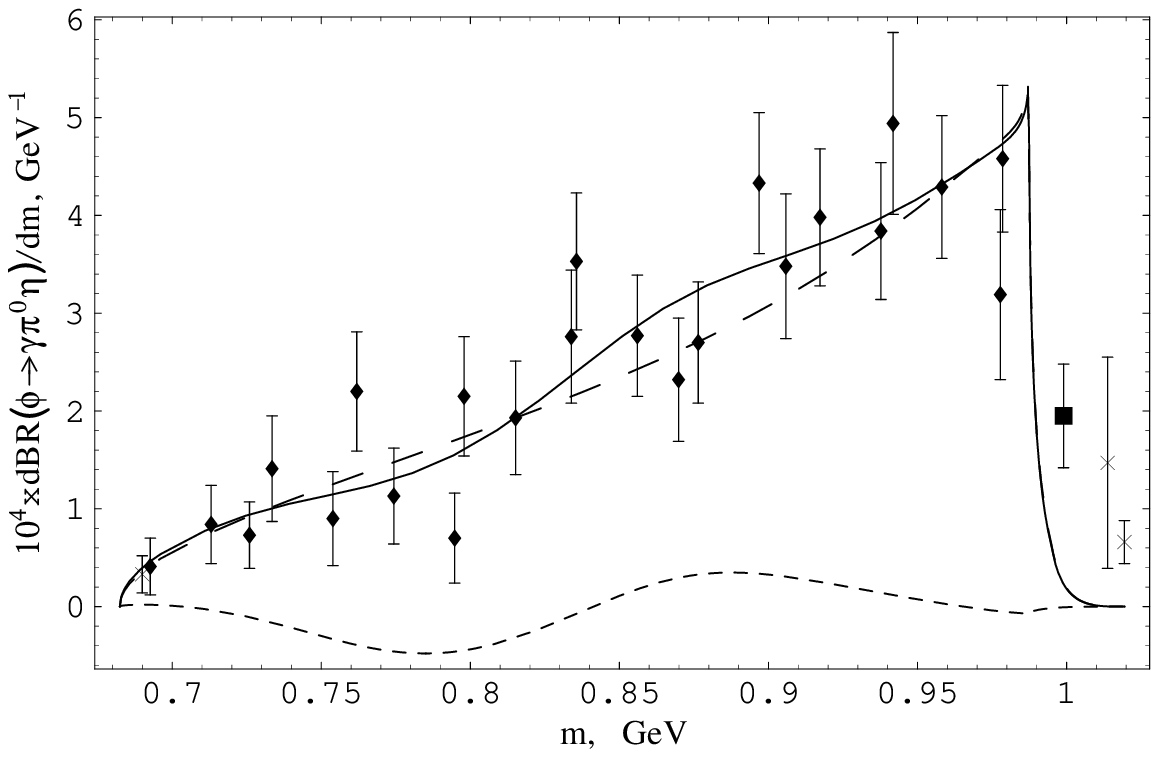} &
\includegraphics[height=1.6in]{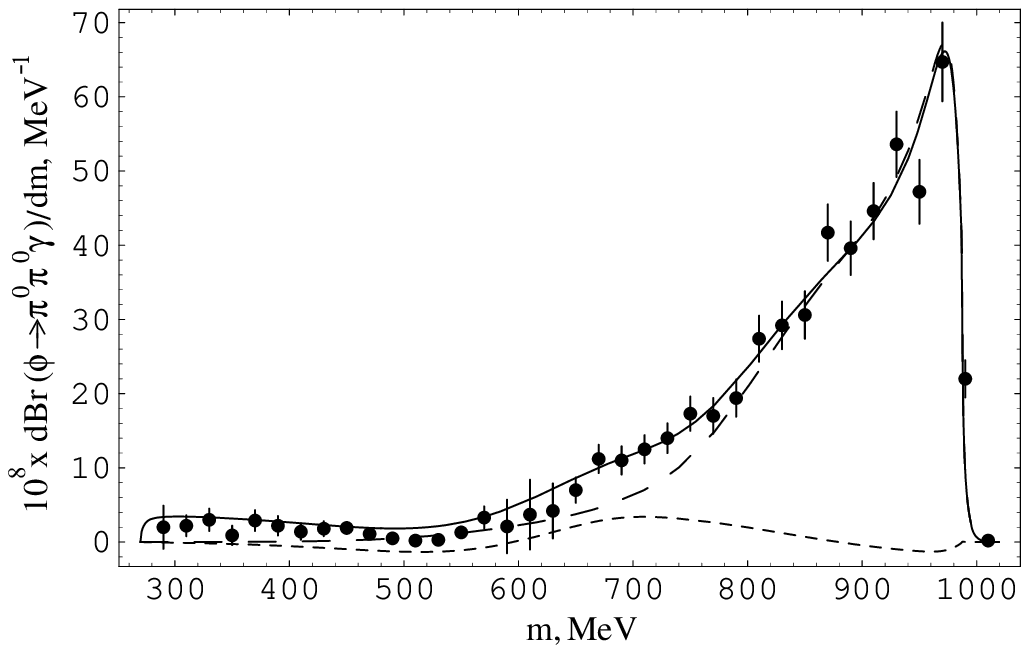}\end{tabular}
\\[0.4cm] {\small Figure 7:}{\footnotesize \ The left (right) plot shows the fit to the
KLOE data for the $\pi^0\eta$ ($\pi^0 \pi^0$) mass spectrum in the
$\phi\to\gamma\pi^0\eta$ ($\phi\to\gamma\pi^0\pi^0$) decay caused by
the $a_0(980)$ ($\sigma(600)+f_0(980)$) production through the
$K^+K^-\ $loop mechanism.}\end{center}

\vspace{4mm} \noindent{\bf\large 15\ \ \boldmath{$K^+K^-$} loop
mechanism is established \cite{A8907}} \vspace{2mm}

In truth this means that $a_0(980)$ and $f_0(980)$ are seen in the
radiative decays of $\phi$ meson owing to  $K^+K^-$ intermediate
state. So, the mechanism of production of $a_0(980)$ and $f_0(980)$
mesons in the $\phi$ radiative decays is established at a physical
level of proof.  We are dealing with the the four-quark transition.
A radiative four-quark transition between two $q\bar q$ states
requires creation and annihilation of an additional $q\bar q$ pair,
i.e., such a transition is forbidden according to the OZI rule,
while a radiative four-quark transition between $q\bar q$ and
$q^2\bar q^2$ states requires only creation of an additional $q\bar
q$ pair, i.e., such a transition is allowed according to the OZI
rule. The large $N_C$ expansion supports this conclusion.

\vspace{4mm} \noindent{\bf\large 16\ \
\boldmath{$a_0(980)/f_0(980)$\,$\to$\,$\gamma\gamma$} \& $q^2\bar
q^2$ model \cite{ADS82,ADS85}} \vspace{2mm}

Twenty seven years ago we predicted the suppression of
$a_0(980)\to\gamma\gamma$ and $f_0(980)\to\gamma\gamma$ in the
$q^2\bar q^2$ MIT model \cite{ADS82},
$$\Gamma(a_0(980)\to\gamma\gamma)\sim\Gamma(f_0(980)\to\gamma\gamma)\sim
0.27\,\mbox{keV}.$$

Experiment supported this prediction $\Gamma
(a_0\to\gamma\gamma)=(0.19\pm 0.07 ^{+0.1}_{-0.07})/B(a_0\to\pi\eta)
\, \mbox{keV, Crystal Ball,}$ $\Gamma (a_0\to\gamma\gamma)=(0.28\pm
0.04\pm 0.1)/B(a_0\to\pi\eta)\, \mbox{keV, JADE,}$ $\Gamma
(f_0\to\gamma\gamma)=(0.31\pm 0.14\pm 0.09)\, \mbox{keV, Crystal
Ball,}$ $\Gamma (f_0\to\gamma\gamma)=(0.24\pm 0.06\pm 0.15)\,
\mbox{keV, MARK II}.$

When in the $q\bar q$ model it was anticipated
$\Gamma(a_0\to\gamma\gamma)=(1.5 - 5.9)\Gamma (a_2\to\gamma\gamma)=
(1.5 - 5.9)(1.04\pm 0.09)\,\mbox{keV,}$
$\Gamma(f_0\to\gamma\gamma)=(1.7 - 5.5)\Gamma (f_2\to\gamma\gamma)=
(1.7 - 5.5)(2.8\pm 0.4)\,\mbox{keV.}$

\vspace{4mm} \noindent{\bf\large 17\ \ Scalar nature and production
mechanisms in \boldmath{$\gamma\gamma$} collisions \cite{AS0509}}
\vspace{2mm}

Recently  the experimental investigations have made great
qualitative advance. The Belle Collaboration  published data on
$\gamma\gamma\to\pi^+\pi^-$ (2007), $\gamma\gamma\to\pi^0\pi^0$
(2008), and $\gamma\gamma\to\pi^0\eta$ (2009), whose statistics are
huge \cite{Be0709}. They not only proved the theoretical
expectations based on the four-quark nature of the light scalar
mesons,  but also have allowed to elucidate the principal mechanisms
of these processes. Specifically, the direct coupling constants of
the $\sigma(600)$, $f_0(980)$, and $a_0(980)$ resonances with the
$\gamma\gamma$ system are small with the result  that their decays
in the two photon are the four-quark transitions caused by the
rescatterings $\sigma$\,$\to $\,$\pi^+\pi^-$\,$\to
$\,$\gamma\gamma$, $f_0(980)$\,$\to $\,$K^+K^-$\,$\to
$\,$\gamma\gamma$ and $a_0(980)$\,$\to $\,$K^+K^-$\,$\to
$\,$\gamma\gamma$ in contrast to the two-photon decays of the
classic $P$ wave tensor $q\bar q$ mesons $a_2(1320)$, $f_2(1270)$
and $f'_2(1525)$, which are caused by the direct two-quark
transitions $q\bar q$\,$\to $\,$\gamma\gamma$ in the main. As a
result the practically model-independent prediction of the $q\bar q$
model $g^2_{f_2\gamma\gamma}:g^2_{a_2\gamma\gamma}=25:9$ agrees with
experiment rather well. The two-photon light scalar widths averaged
over resonance mass distributions $\langle\Gamma_{f_0\to\gamma
\gamma}\rangle_{\pi\pi}$\,$\approx$\,0.19 keV,
$\langle\Gamma_{a_0\to\gamma \gamma}\rangle_{\pi\eta}$\,$\approx
$\,0.34 keV and $\langle\Gamma_{\sigma
\to\gamma\gamma}\rangle_{\pi\pi}$\,$\approx$\,0.45 keV. As to the
ideal $q\bar q$ model prediction
$g^2_{f_0\gamma\gamma}:g^2_{a_0\gamma\gamma}=25:9$, it is excluded
by experiment.

\vspace{4mm} \noindent{\bf\large 18\ \ Dynamics of \boldmath
$\gamma\gamma\to\pi^+\pi^-$, $\gamma\gamma\to\pi^0\pi^0$ and
$\gamma\gamma\to\pi^0\eta$ \cite{AS0509}} \vspace{2mm}

The following figures give a scetch of our treatment of the Belle
data on the reactions $\gamma\gamma\to\pi\pi$ and
$\gamma\gamma\to\pi^0\eta$.

\begin{center}\begin{tabular}{cc}
\includegraphics[width=7.1cm]{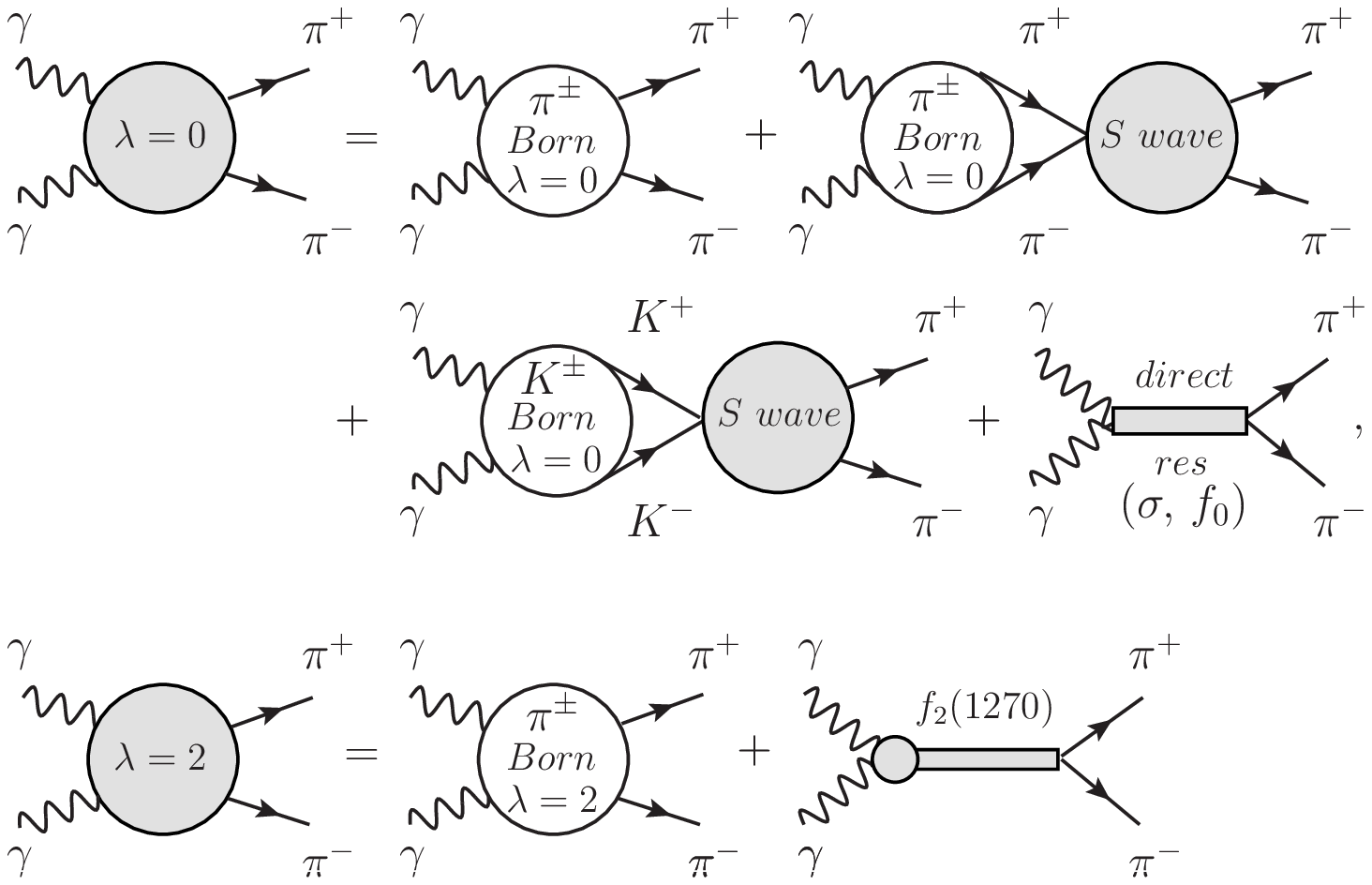}\  & \ \ \ \includegraphics[width=7.1cm]{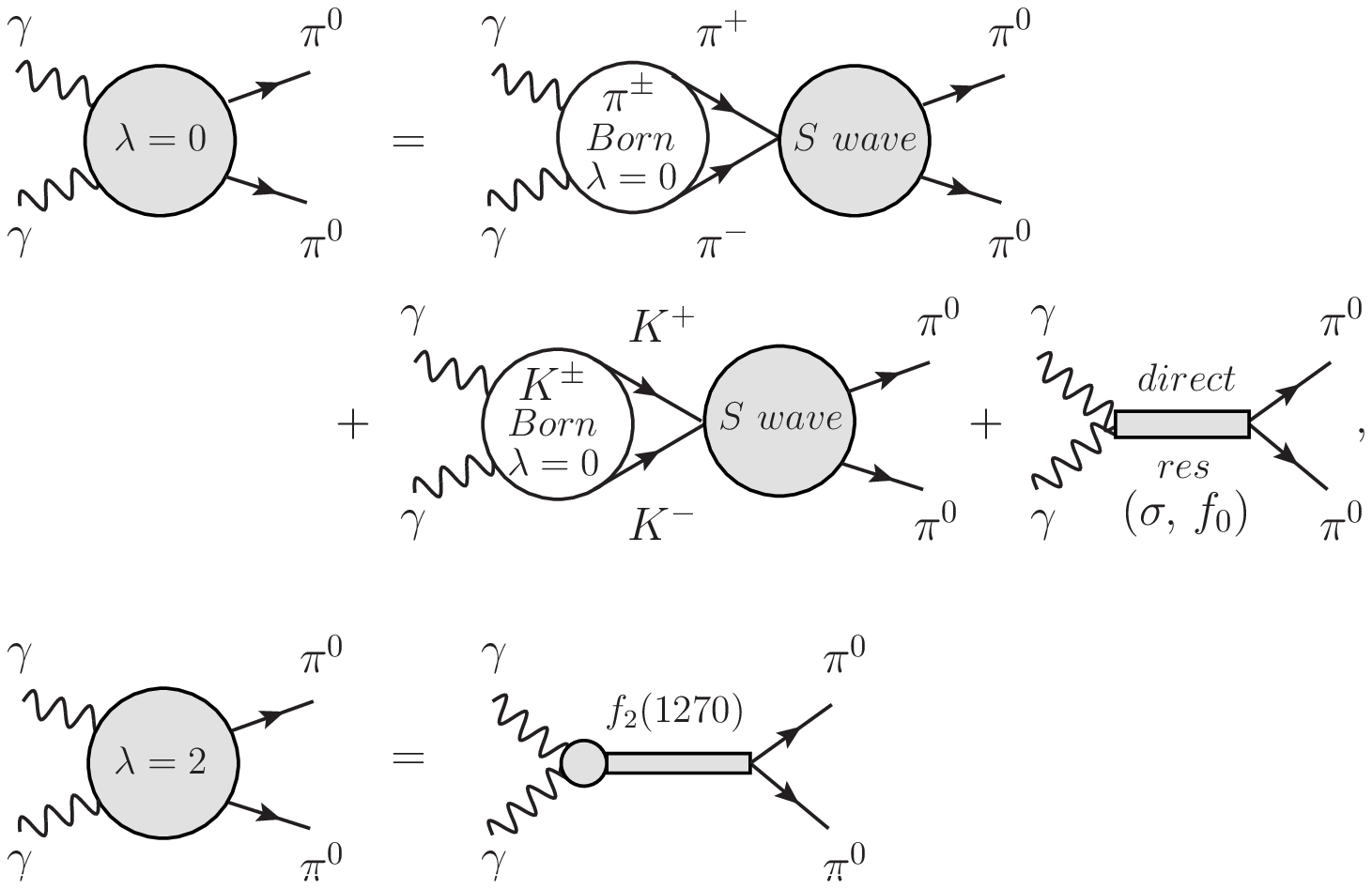}
\\[0.2cm]
\small{Figure 8:}{\footnotesize \ Diagrammatical representation}\  &
\ \small{Figure 9:}{\footnotesize \ Diagrammatical representation}
\\ {\footnotesize  for the helicity amplitudes
$\gamma\gamma\to\pi^+\pi^-$.}\  & \  {\footnotesize  for the
helicity amplitudes $\gamma\gamma\to\pi^0\pi^0$.}
\end{tabular}\end{center}

\begin{center}
\includegraphics[width=7.1cm]{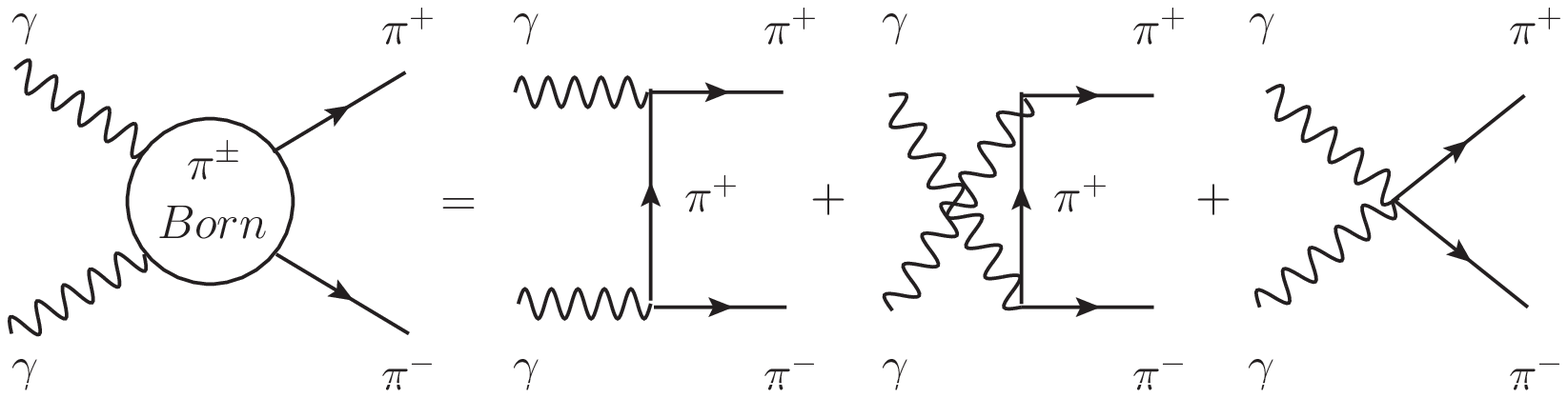}
\\
\includegraphics[width=7.1cm]{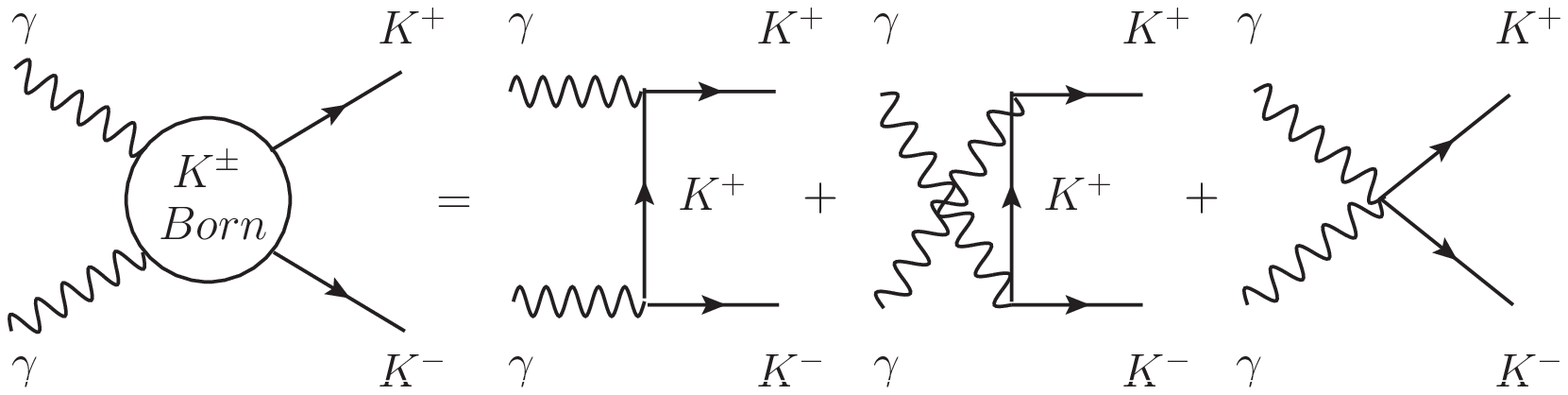}
\\[0.2cm]
\small{Figure 10:}{\footnotesize \ The Born sources for
$\gamma\gamma\to\pi^+\pi^-$ and $\gamma\gamma\to
K^+K^-$.}\end{center}

\begin{center}\vspace{0.1cm}
\includegraphics[width=5.1cm]{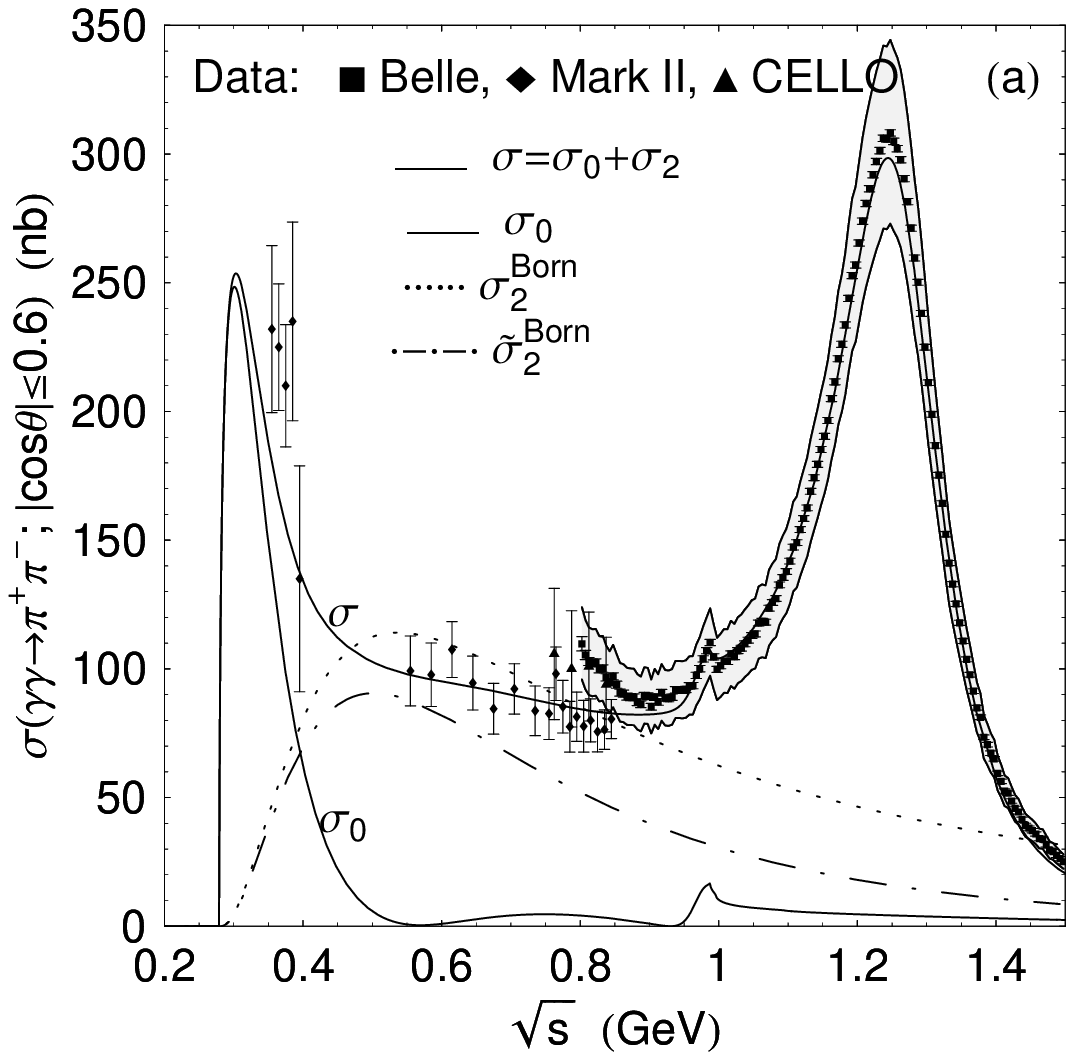}
\includegraphics[width=5.1cm]{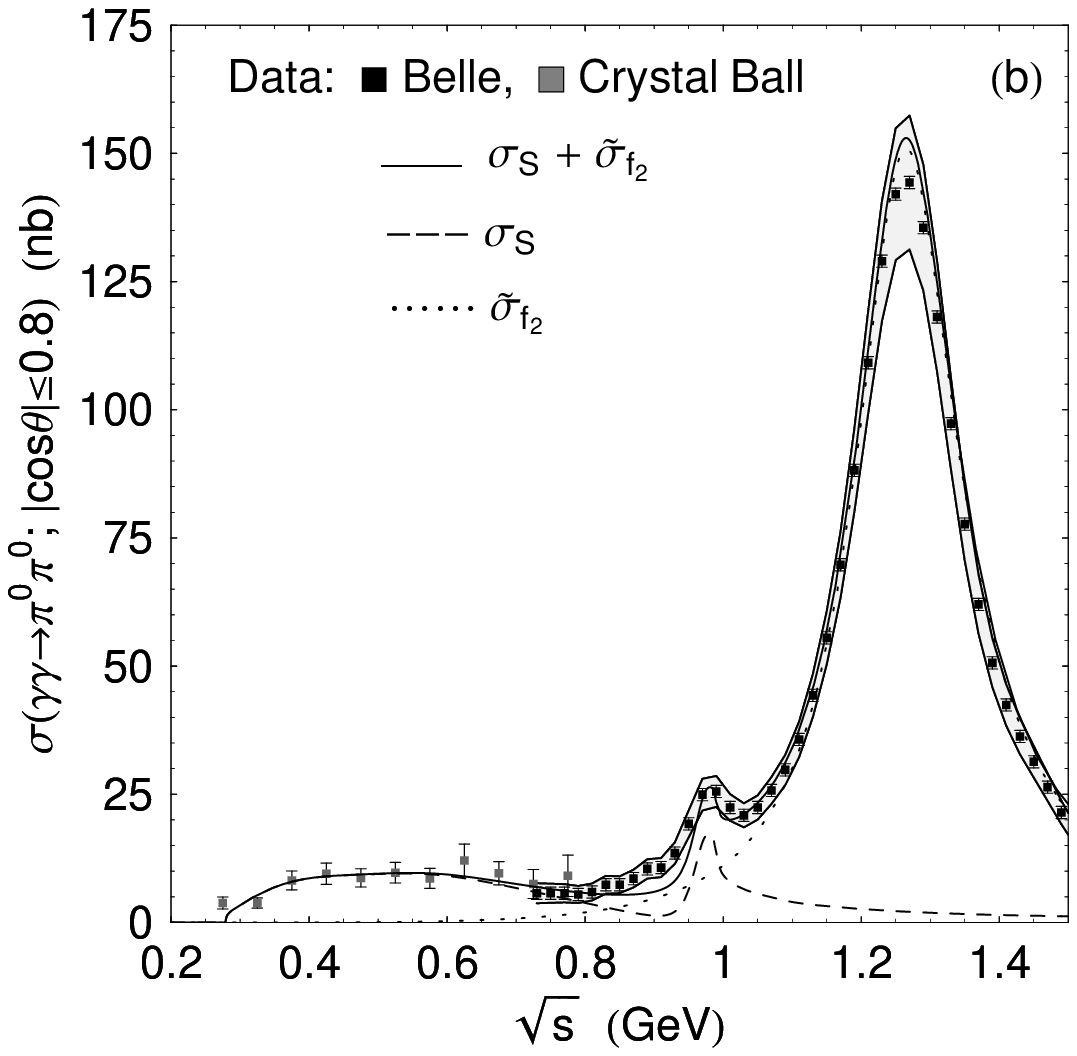}
\\[0.2cm]
\small{Figure 11:}{\footnotesize \ The descriptions of the Belle
data on $\gamma\gamma\to\pi^+\pi^-$ (a) and on
$\gamma\gamma\to\pi^0\pi^0$ (b).}\end{center}

\begin{center}\begin{tabular}{cc}
\includegraphics[width=7.1cm]{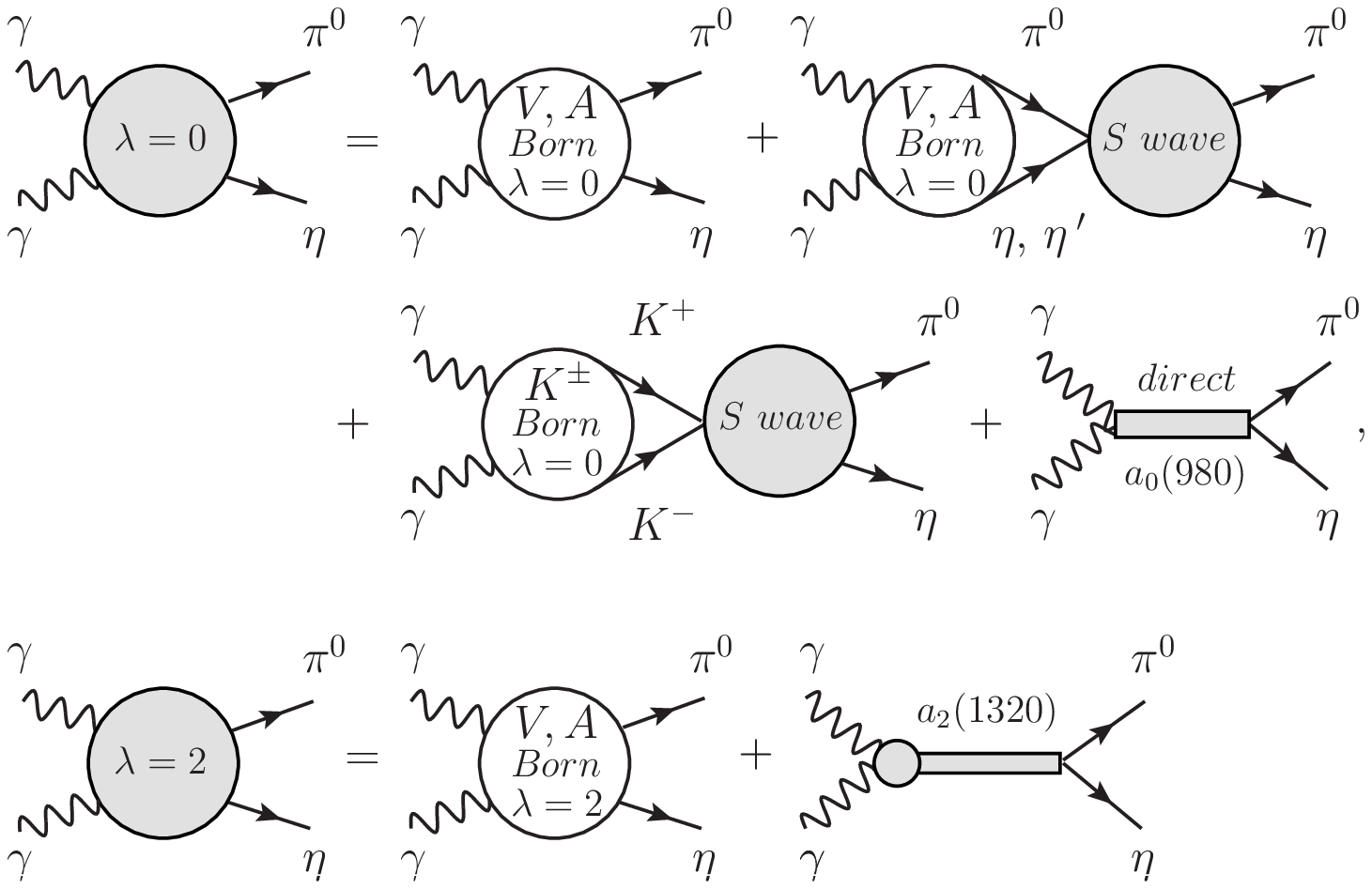}\  & \ \ \
\raisebox{12.0mm}{$\includegraphics[width=7.1cm]{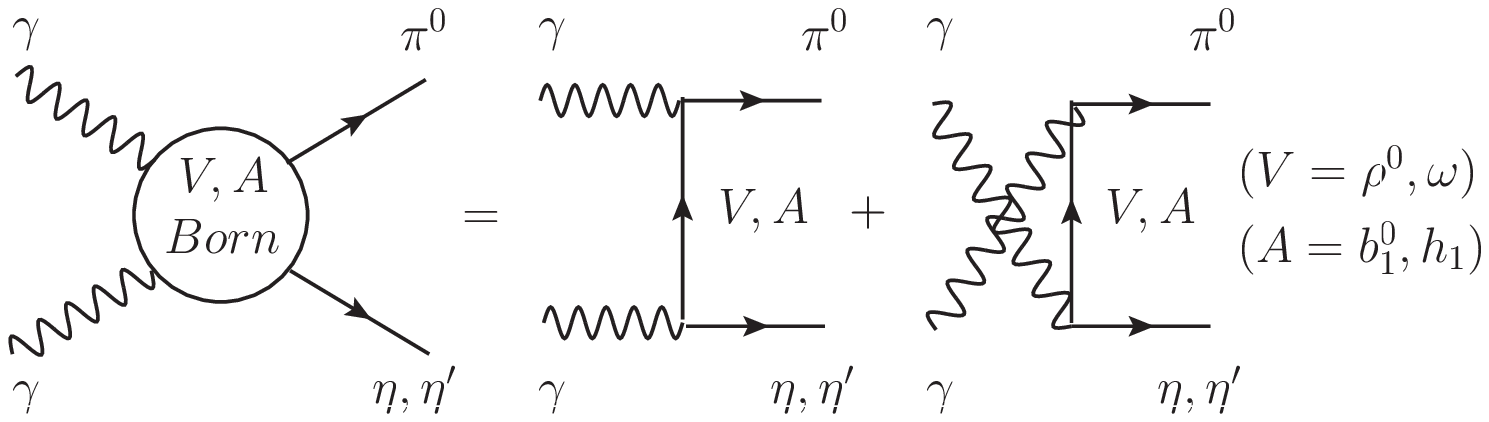}$}
\\[0.2cm]
\small{Figure 8:}{\footnotesize \ Diagrammatical representation}\  &
\ \small{Figure 9:}{\footnotesize \ The Born sources for
$\gamma\gamma\to\pi^0\eta$.}
\\ {\footnotesize  for the helicity amplitudes
$\gamma\gamma\to\pi^0\eta$.}\  &
\end{tabular}\end{center}

\begin{center}
\includegraphics[width=5.1cm]{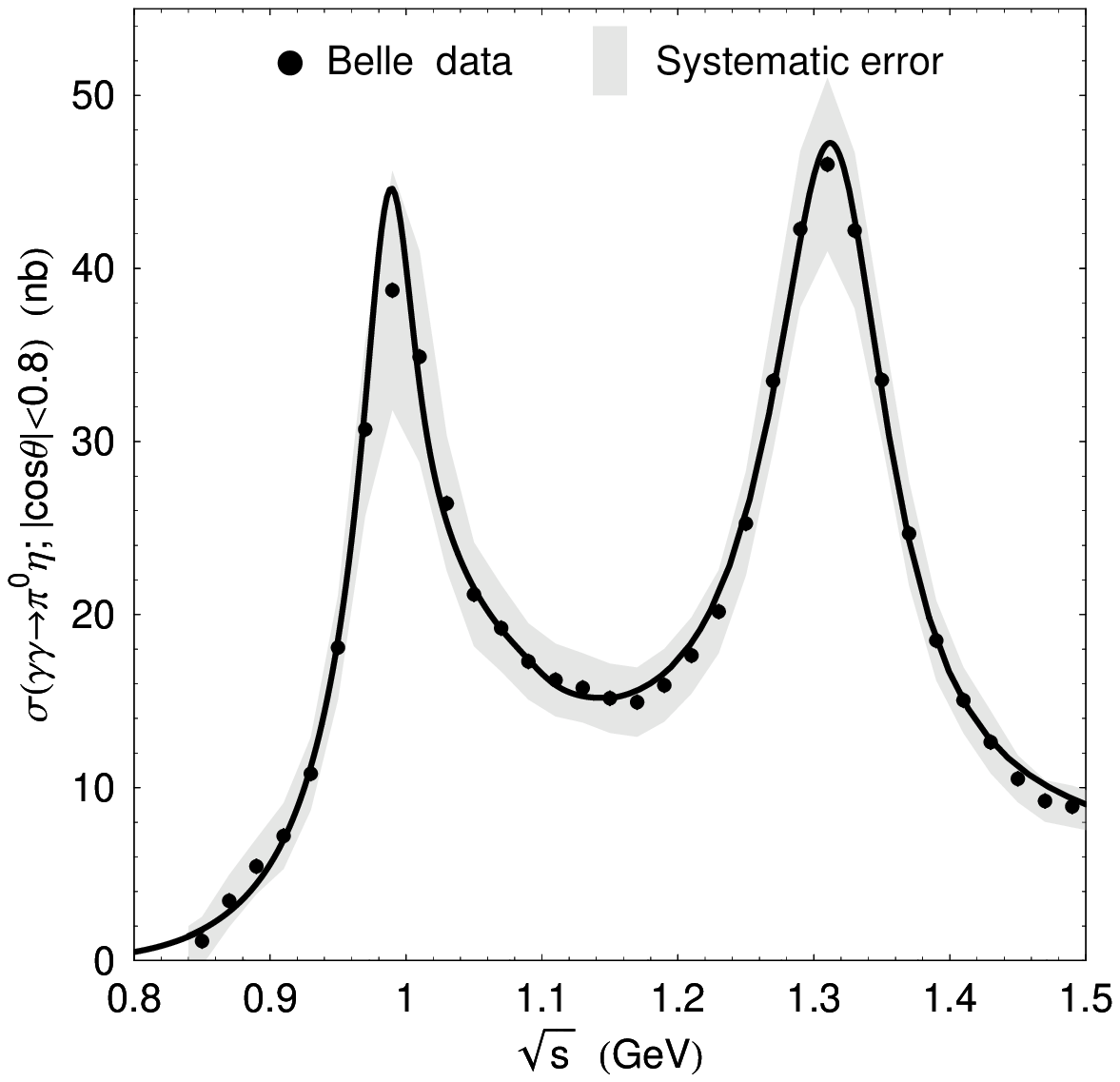}
\\
\small{Figure 14:}{\footnotesize \ The description of the Belle data
on $\gamma\gamma\to\pi^0\eta$.}\end{center}

\begin{center}
\includegraphics[width=10cm]{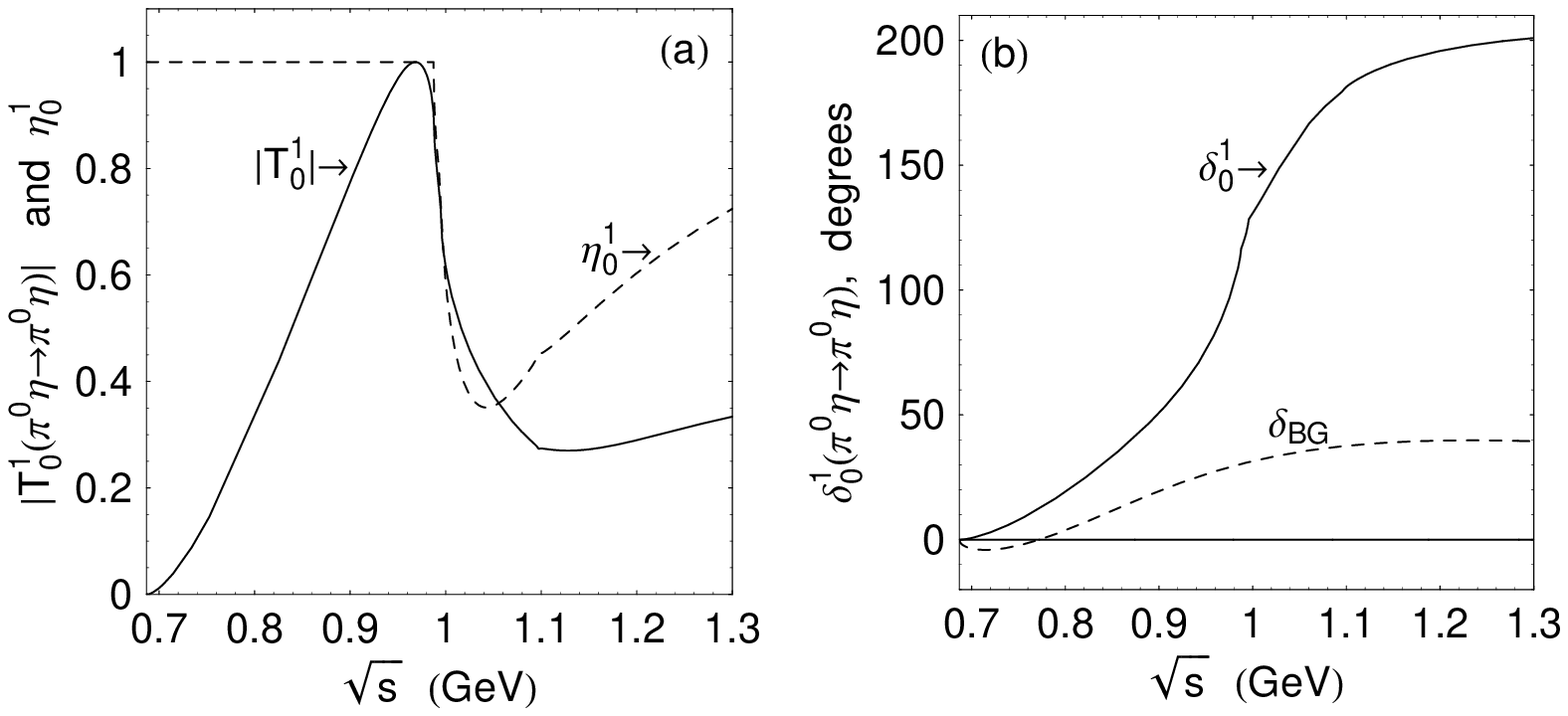}
\\
\small{Figure 15:}{\footnotesize \ Preliminary results for the
$\pi^0\eta\to\pi^0\eta$ reaction amplitude.}
\end{center}

\vspace{4mm} \noindent{\bf\large 19\ \ Summary} \vspace{2mm}

The mass  spectrum of the light scalars, $\sigma (600)$, $\kappa
(800)$, $f_0(980)$, $a_0(980)$, gives an idea of their $q^2\bar q^2$
structure.

Both intensity and mechanism of the $a_0(980)/f_0(980)$ production
in the radiative decays of $\phi(1020)$, the  $q^2\bar q^2$
transitions $\phi\to K^+K^-\to\gamma[a_0(980)/f_0(980)]$, indicate
their $q^2\bar q^2$ nature.

Both intensity and mechanism of the scalar meson decays into $\gamma
\gamma$, the $q^2\bar q^2$ transitions,
$\sigma(600)\to\pi^+\pi^-\to\gamma\gamma$, $f_0(980)/a_0(980)\to
K^+K^-\to\gamma\gamma$, indicate their $q^2\bar q^2$ nature also.

In addition, the absence of $J/\psi$ $\to\gamma f_0(980)$,
$a_0(980)\rho$, $f_0(980)\omega$ in contrast to the intensive
$J/\psi$ $\to$ $\gamma f_2(1270)$, $\gamma f'_2(1525)$,
$a_2(1320)\rho$, $f_2(1270)\omega$ decays intrigues against the $P$
wave $q\bar q$ structure of $a_0(980)$ and $f_0(980)$ also.

\vspace{4mm}  This work was supported in part by the RFFI Grant No.
07-02-00093 from the Russian Foundation for Basic Research and by
the Presidential Grant No. NSh-1027.2008.2 for Leading Scientific
Schools.

\end{document}